\documentclass[twocolumn]{aastex631}

\usepackage{booktabs}
\usepackage{multirow}

\received{11 December 2024}
\revised{12 February 2025}
\accepted{15 February 2025 -- ApJL}

\begin{document}

\title{A Time-Resolved High-Resolution Spectroscopic Analysis of Ionized Calcium and\\Dynamical Processes in the Ultra-Hot Jupiter HAT-P-70~b}
\shorttitle{Time-Resolved Transmission Spectroscopy of HAT-P-70~b}
\shortauthors{Langeveld et al.}

\correspondingauthor{Adam Langeveld}
\email{alangev1@jh.edu}

\author[0000-0002-4451-1705]{Adam B. Langeveld}
\affiliation{Department of Physics and Astronomy, Johns Hopkins University,  Baltimore, MD 21218, USA}
\affiliation{Department of Astronomy and Carl Sagan Institute, Cornell University, Ithaca, NY 14850, USA}

\author[0000-0001-9796-2158]{Emily K. Deibert}
\affiliation{International Gemini Observatory/NSF NOIRLab, Casilla 603, La Serena, Chile}

\author[0000-0003-0672-7123]{Mitchell E. Young}
\affiliation{Astrophysics Research Centre, Queen’s University Belfast, Belfast, BT7 1NN, UK}

\author[0000-0001-6391-9266]{Ernst de Mooij}
\affiliation{Astrophysics Research Centre, Queen’s University Belfast, Belfast, BT7 1NN, UK}

\author[0000-0001-5349-6853]{Ray Jayawardhana}
\affiliation{Department of Physics and Astronomy, Johns Hopkins University,  Baltimore, MD 21218, USA}

\author[0000-0001-8589-4055]{Chris Simpson}
\affiliation{Gemini Observatory, NSF's NOIRLab, 670 N A`ohōkū Place, Hilo, HI 96720, USA}

\author[0000-0001-7836-1787]{Jake D. Turner}
\affiliation{Department of Astronomy and Carl Sagan Institute, Cornell University, Ithaca, NY 14850, USA}

\author[0000-0001-6362-0571]{Laura Flagg}
\affiliation{Department of Physics and Astronomy, Johns Hopkins University,  Baltimore, MD 21218, USA}



\begin{abstract}
We present the first transmission spectroscopy study of an exoplanet atmosphere with the high-resolution mode of the new Gemini High-resolution Optical SpecTrograph (GHOST) instrument at the Gemini South Observatory. We observed one transit of HAT-P-70~b -- an ultra-hot Jupiter with an inflated radius -- and made a new detection of the infrared Ca\,{\sc ii} triplet in its transmission spectrum. The depth of the strongest line implies that a substantial amount of Ca\,{\sc ii} extends to at least 47\,\% above the bulk planetary radius. The triplet lines are blueshifted between $\sim$~3 to 5~km\,s$^{-1}$, indicative of strong dayside-to-nightside winds common on highly irradiated gas giants. Comparing the transmission spectrum with atmospheric models that incorporate non-local thermodynamic equilibrium effects suggests that the planetary mass is likely between 1 to 2~$M_{\rm J}$, much lighter than the upper limit previously derived from radial velocity measurements. Importantly, thanks to the the high signal-to-noise ratio achieved by GHOST/Gemini South, we are able to measure the temporal variation of these signals. Absorption depths and velocity offsets of the individual Ca\,{\sc ii} lines remain mostly consistent across the transit, except for the egress phases, where weaker absorption and stronger blueshifts are observed, highlighting the atmospheric processes within the trailing limb alone. Our study demonstrates the ability of GHOST to make time-resolved detections of individual spectral lines, providing valuable insights into the 3D nature of exoplanet atmospheres by probing different planetary longitudes as the tidally locked planet rotates during the transit.
\end{abstract}



\section{Introduction}
\label{sec:introduction}

The Gemini High-resolution Optical SpecTrograph (GHOST) is the new flagship high-resolution spectrograph at the Gemini South Observatory, Cerro Pachón, Chile \citep{ireland2012, ireland2014, mcconnachie2024, kalari2024}. 
While designed to accomplish several key science objectives, it holds particular promise to deliver state-of-the-art optical spectra for in-depth characterization of transiting exoplanet atmospheres, both with transmission spectroscopy during the primary transit and emission spectroscopy at orbital phases around the secondary eclipse. GHOST's potential for such studies was demonstrated during the System Verification run in May 2023: with only $\sim3$ hours of observations during the post-eclipse orbital phases of the ultra-hot Jupiter (UHJ) WASP-189~b, neutral iron was detected at a significance of 7.5\,$\sigma$, confirming the presence of a thermal inversion in the planet's dayside atmosphere \citep{deibert2024}.

UHJs are typically tidally locked with short-period orbits in close proximity to hot (A- or F-type) host stars. Their permanent daysides receive intense stellar irradiation, causing temperatures to exceed $\sim$~2200~K and creating large day-to-night temperature gradients \citep{arcangeli2018, bell2018, parmentier2018}. Observations and theoretical models have shown that these extreme conditions lead to exotic atmospheres composed of a diverse array of gaseous metallic atoms and ions \citep[e.g.][]{kitzmann2018, tabernero2021, borsa2021, borsato2023, pelletier2023, prinoth2024}, together with strong winds and atmospheric flow from the dayside to the nightside \citep[e.g.][]{showman2008, showman2020, rauscher2014, heng2015, tan2019, ehrenreich2020, seidel2021}.
High-resolution spectroscopy has proven to be a powerful probe of UHJ atmospheres since it can resolve densely packed, narrow features that are imprinted within their transmission spectra due to absorption from chemical species. In particular, spectral features from the commonly detected refractory metal atoms and ions in UHJs are most prevalent at optical wavelengths.

``Time-resolved'' high-resolution transmission spectroscopy (where a significant atmospheric signal is detected at multiple phases across the transit) is an innovative technique that offers unprecedented detail into 3D atmospheric composition and dynamics, but has typically only been feasible using cross-correlation focused methods \citep[e.g.][]{ehrenreich2020, kesseli2021, gandhi2022, pelletier2023, prinoth2023, wardenier2024, simonnin2024}. With this approach, hundreds of absorption lines from a particular chemical species are effectively averaged over the entire spectral range, significantly improving the signal-to-noise ratio (SNR). Such detections have led to the discovery of interesting phenomena with a temporal and spatial dependence, such as asymmetries in the atmospheric signals between the morning and evening terminators that could be explained by a multitude of physical mechanisms \citep[e.g.][]{ehrenreich2020, wardenier2021, savel2022, savel2023, beltz2023, nortmann2025}.
Performing the same time-resolved studies for individual absorption lines (without cross-correlating) is notably more difficult, because the spectra often need to be combined over the whole transit to make significant narrowband detections. However, pioneering efforts using the latest state-of-the-art spectrographs at 8--10\,m telescopes have shown recent success \citep[e.g.][]{seidel2023, prinoth2024}.

GHOST's high-resolution mode has a resolving power of \mbox{$R\sim76\,000$} and is able to obtain useful (throughput $>2$\%) flux over a wavelength range of $\sim$~3830--10\,000\,{\AA} \citep{kalari2024}, making it an ideal instrument for characterizing UHJ atmospheres. Further information about the design and specifications of GHOST, including overviews of both the ``standard-resolution'' and ``high-resolution'' modes, are presented by \citet{mcconnachie2024} and \citet{kalari2024}.

During the GHOST Shared Risk run in December 2023, we observed one transit of the UHJ \mbox{HAT-P-70~b} to probe the terminator with high-resolution spectroscopy and investigate the chemical and dynamical processes across the boundary between the dayside and the nightside.
\mbox{HAT-P-70~b} has an equilibrium temperature of $\sim$~2550~K, well above the $\sim$~2200~K transition between the ``hot'' and ``ultra-hot'' regimes. It resides on a near-polar orbit \mbox{($P=2.74$~days)} around a bright ($m_{\mathrm V} = 9.5$) and rapidly rotating ($v\sin i=99.85$~km\,s$^{-1}$) A-type star, with a transit duration of 3.48~hours \citep{zhou2019} -- a full list of system parameters can be found in Table~\ref{tab:parameters}. HAT-P-70~b is particularly conducive to transmission spectroscopy since it is one of the most inflated UHJs with a radius of 1.87~$R_\mathrm{J}$, and has a large transit depth ($> 1$\%), therefore acting as a great test case for this first transmission spectroscopy study with the high-resolution mode of GHOST.

One transit of HAT-P-70~b was previously observed with the High Accuracy Radial velocity Planet Searcher in the Northern hemisphere (HARPS-N) at the Telescopio Nazionale Galileo (TNG). Absorption features from several chemical species in the transmission spectrum were detected using the cross-correlation technique, and individual measurements were made for the strong optical absorption lines of hydrogen (H$\gamma$, H$\beta$, H$\alpha$), sodium (Na\,{\sc i} doublet), and ionized calcium (Ca\,{\sc ii} H and K lines) \citep{bello-arufe2022}.

In this paper, we present new time-resolved detections of the infrared Ca\,{\sc ii} triplet lines, taking advantage of the wider wavelength coverage of GHOST and the increased SNR provided by the 8.1\,m mirror of Gemini~South. A full survey of chemical species in the transmission spectrum of HAT-P-70~b using these data will follow in the future. \mbox{Information} about the observations can be found in Section~\ref{sec:observations}, and Section~\ref{sec:methods} outlines the transmission spectrum extraction. In Section~\ref{sec:results} we present the detection of the Ca\,{\sc ii} triplet and discuss inferences that can be made about the upper atmosphere of HAT-P-70~b from both the combined and time-resolved transmission spectra, as well as comparison with atmospheric models. Our conclusion and outlook for future work can be found in Section~\ref{sec:conclusions}.

\section{Observations and Data Reduction}
\label{sec:observations}

We observed one transit of HAT-P-70~b with GHOST/Gemini~South on 2023 December 15 (Program GS-2023B-FT-105, PI: Langeveld) during the Shared Risk run. With the ``high-resolution'' mode, the target and the background sky were observed simultaneously.

\begin{figure}
    \centering
    \includegraphics[width=\columnwidth]{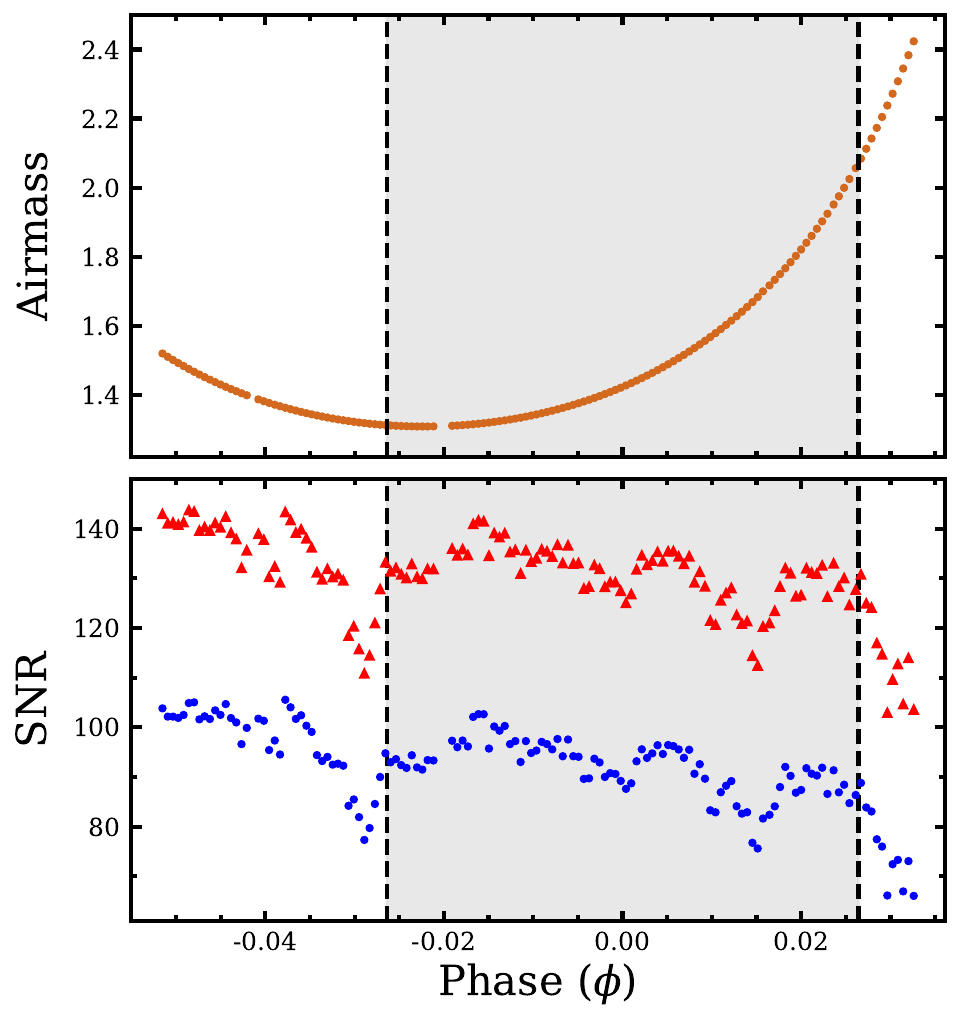}
    \vskip-1em
    \caption{\textit{Top panel:} Change in airmass throughout the observations. One exposure was discarded during the transit. There was a temporary loss of guiding during the pre-transit sequence. \textit{Bottom panel:} Average SNR across the full spectrum for each blue and red exposure (marked with blue circles and red triangles respectively). The black dashed lines and grey shaded region highlight the in-transit phases.}
    \label{fig:airmass_snr}
    \vskip-0.2em
\end{figure}

GHOST has two arms (blue and red) -- the full spectral range is divided at $\sim5300$\,{\AA} with a beam splitter, before the cross dispersion. Exposures from both the blue and red detectors were configured to start simultaneously to ensure that the same orbital phases were combined and analyzed together.
Medium read modes were used for both detectors, with $1\times4$ (spectral $\times$ spatial) binning to significantly decrease the readout times whilst maintaining the full spectral resolution. 
140 pairs of exposures were acquired over $\sim5.5$~hr, each with a 120\,s exposure time -- this consisted of $\sim3.5$~hr for the transit and $\sim2$~hr for the pre- and post-transit baseline. We aimed to achieve a SNR~$\gtrsim100$ over the majority of the spectral range whilst avoiding Doppler smearing of the planetary signal \citep{boldt-christmas2024}. Figure~\ref{fig:airmass_snr} shows the airmass variation and the average SNR for the blue and red arms throughout the sequence. One pair of exposures starting on 2023 December 16 at 04:10:03 (UT) was discarded due to a software restart after not acquiring slit images.
Of the 139 remaining exposures, 87 were obtained during the in-transit phases, leaving 52 to build the high SNR out-of-transit baseline.

\subsection{Data Reduction and Spectrograph Focusing}
\label{sec:focus}

During the GHOST Shared Risk run, the spectrograph suffered from an issue where the focus was offset by $\sim500$\,{\micron} ($\sim4$\% different from the nominal value) for the red and blue science cameras. This resulted in an increase in inter-order flux, a loss of resolution, and artefacts appearing as periodic ``ripples'' visible in the extracted spectra. 
This was only a temporary issue experienced during the GHOST Shared Risk run; newer data acquired since then no longer exhibit the same problems as described here.

The data were reduced with version 1.1.1 of the GHOST Data Reduction (\texttt{GHOSTDR}) software \citep{ireland2018, hayes2022} utilizing the Data Reduction for Astronomy from Gemini Observatory North and South (\texttt{DRAGONS}) platform \citep{labrie2023}. \texttt{GHOSTDR} v1.1.1 has an improved reduction algorithm designed specifically to account for the unfocused Shared Risk data. Namely, it Gaussian smooths the slit profile from the slit-viewer camera (which was in focus) to more accurately represent the profile seen by the spectrograph, and uses different regions of the slit to extract the object and sky spectra. This decreased the magnitude of the spurious artefacts in the extracted spectra (compared to \texttt{GHOSTDR} v1.1.0), retaining them only at a 1--2\% level. The location of these artefacts did not change significantly throughout the night, causing minimal impact on our analysis of the \mbox{extracted exoplanetary spectra (see Section~\ref{sec:methods}).}

The \texttt{DRAGONS/GHOSTDR} pipeline takes the blue and red detector images and performs the bias subtraction, bad pixel masking, and flat-fielding. Then, the observed spectrum is optimally extracted with cosmic-ray rejection (including optional sky subtraction), and a wavelength solution and optional barycentric correction are applied, to produce outputs in both 2D (a spectrum for each order) and 1D (a single spectrum with all orders merged) form, resulting in two data products for each arm. We opted to reduce the data with the sky subtraction and barycentric correction turned off since (a) no significant features were visible in the sky spectrum in the region of the Ca\,{\sc ii} triplet, and (b) we require the spectra to be in the Earth rest frame for accurate telluric correction (see Section~\ref{sec:methods}).

The impact of the unfocused spectrograph is still being assessed; we hypothesize that it could result in the planetary atmospheric signals being slightly broadened, or imperfect corrections to the ripples causing the line shapes to be distorted.
Therefore, in this study, we choose to demonstrate GHOST's capability for time-resolved transmission spectroscopy of individual absorption lines for only one chemical species (the Ca\,{\sc ii} triplet). Following a robust assessment of the focusing implications, a full survey of the atmospheric chemistry of \mbox{HAT-P-70~b} will follow in future work.

\section{Transmission Spectroscopy}
\label{sec:methods}

We took the 1D data products from the \texttt{DRAGONS/GHOSTDR} pipeline and followed the methods described in \citet{langeveld2022} to clean and process the data and extract the transmission spectrum of HAT-P-70~b. Minor changes to the pipeline are outlined below. The blue and red arms of GHOST were treated independently using the same methods, except when generating the telluric models (see below).

Telluric contamination was removed by generating models of the transmission through Earth’s atmosphere with \texttt{molecfit} version~4.3.1 \citep{smette2015, kausch2015}, which has been applied successfully for many previous studies of exoplanet atmospheres with high-resolution spectroscopy \citep[e.g.][]{allart2017, langeveld2021}. Since the blue arm of GHOST covers wavelengths relatively devoid of strong telluric lines, the blue and red spectra were joined together to produce one telluric model per exposed phase; the data were cut at 5350\,{\AA} and the blue spectra were scaled to the same level as the red by matching the medians of the last 500 blue flux values to the first 500 red flux values. This is justified given that both detectors were configured to start and end each exposure simultaneously, thus covering the same time interval and airmass range. The resulting \texttt{molecfit} model was interpolated back onto the original wavelength grid of the separate blue/red spectra and divided out to remove the telluric contamination. The region encapsulating the Ca\,{\sc ii} triplet only contains weak tellurics ($\gtrsim0.95$ transmission); thus, for this study, there are no issues with the model failing to provide an accurate correction for very deep lines.

We accounted for Doppler shifts due to Earth's barycentric radial velocity and the systemic velocity, but not the stellar radial velocity due to the uncertainty in the mass of HAT-P-70~b and the fact that it is a rapidly rotating star with broad spectral lines. The stellar components of the spectra were then removed by dividing by a master out-of-transit spectrum (the weighted average of all out-of-transit spectra, with each flux value assigned a weight of $1/\sigma^2$), resulting in ``residual spectra'' – the in-transit residuals contain the exoplanet's atmospheric transmission features (plus noise) and the out-of-transit residuals represent only the noise.

\begin{figure*}
    \centering
    \includegraphics[width=\textwidth]{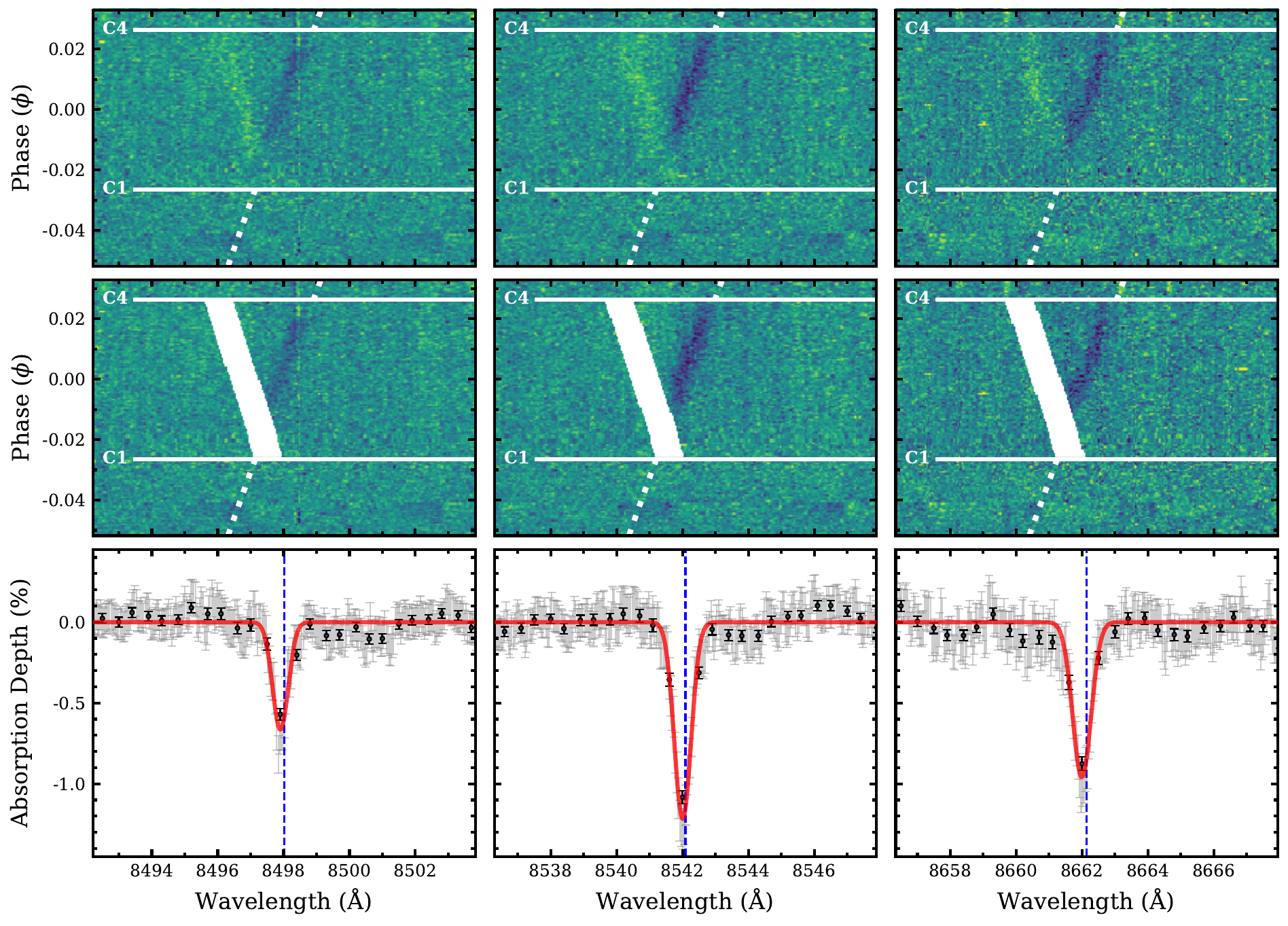}
    \caption{
    \textit{Top row:} Residual spectra around the Ca\,{\sc ii} triplet lines after removing the telluric and stellar components, and dividing each observed spectrum by the master out-of-transit spectrum. The first (C1) and fourth (C4) transit contact points are marked with horizontal white lines, and the expected trails of the planetary atmospheric signals are highlighted by the white dotted lines (which are diagonal due to the Doppler motion of the planet). Atmospheric absorption from Ca\,{\sc ii} is clearly visible from the dark diagonal features following the projections of the expected planetary trails. The spurious features caused by the RM effect are visible as bright diagonal trails in the opposite direction.
    \textit{Middle row:} Same as the top row, with the dominant RM trail masked out. The residual spectra now only contain the planetary signals.
    \textit{Bottom row:} Combined transmission spectrum calculated using the weighted average of the in-transit data in the middle panel after Doppler shifting into the planetary rest frame. The data are displayed in their full resolution (grey) and binned 10$\times$ (black). Gaussian fits to the Ca\,{\sc ii} absorption lines are shown in red, and the blue dashed lines mark the expected position of the absorption lines in the planetary rest frame.}
    \vspace{2em}
    \label{fig:transmission_spectrum}
\end{figure*}

Individual absorption lines from various chemical species are typically too weak to be visible over the noise in the residual spectra, but they should follow a sinusoidal path due to the orbital motion of the planet -- during the transit, this path is a diagonal line between the phases of the first and fourth transit contact points. However, thanks to the high SNR achieved by GHOST/Gemini~South, strong features such as the Ca\,{\sc ii} triplet lines are remarkably visible across the whole time series, as shown in Figure~\ref{fig:transmission_spectrum}. Such clear ``time-resolved'' detections of single lines have only become achievable on a wide scale in recent years with 8--10\,m class telescopes and state-of-the-art high-resolution spectrographs \citep[e.g. \mbox{ESPRESSO/VLT} and \mbox{MAROON-X}/Gemini~North:][]{seidel2023, prinoth2024}.
Therefore, this demonstration places GHOST among the few instruments that can achieve the required SNR to conduct such studies and analyze how the absorption features vary in time.

The top row of Figure~\ref{fig:transmission_spectrum} also shows spurious signals imprinted within the transmission spectra due to the Rossiter-McLaughlin (RM) effect, which follow a diagonal path in the opposite direction to the planetary trail \citep[for further discussion, see, e.g.][]{rossiter1924, mclaughlin1924, gaudi2007, triaud2018, prinoth2024}. These signals can result in false detections of atmospheric species \citep[e.g.][]{casasayas-barris2020, casasayas-barris2021}, so must be removed to fully assess the extent of absorption in the planetary atmosphere.
As seen in Figure~\ref{fig:transmission_spectrum} (top row), the RM signals have a similar amplitude as those from the planetary atmosphere, evident by the overlapping region near the transit ingress where the RM and atmospheric features essentially cancel each other out. We modelled the RM effect and centre-to-limb variations (CLV) using the same grid-based approach as in \citet{turner2020} and \citet{deibert2021}. However, given the large extent of the planet in the Ca\,{\sc ii} lines, which is not accounted for in this model, there is a strong residual signal during the phase range where the planet's signal overlaps with the RM signal. Correcting for this would require a model that simultaneously accounts for the planet's atmosphere and the wavelength-dependent RM effect \citep[e.g.][]{dethier2023}, which is beyond the scope of this work. We therefore masked out this region, following some other studies that have measured time-resolved signals \citep[e.g.][]{borsa2021, kesseli2022, simonnin2024}, as shown in the second row of Figure~\ref{fig:transmission_spectrum}.

Finally, each spectrum was Doppler shifted into the planetary rest frame using radial velocity values calculated from the parameters listed in Table~\ref{tab:parameters}, and subtracted by 1 \citep[e.g.][]{wyttenbach2015, langeveld2022}. Then, all fully in-transit phases (where at least some of the planet is transiting for the entire exposure duration) were combined with a weighted average ($\text{weights}=1/\sigma^2$ for all flux values). The result is a single high SNR transmission spectrum that represents the average atmospheric absorption over the whole transit, shown in Figure~\ref{fig:transmission_spectrum} (bottom panel).

We note that there are some residual patterns visible in the top panels of Figure~\ref{fig:transmission_spectrum}, between phases of -0.05 and -0.04, which likely originate from imperfect corrections of the ripples in the out-of-focus data. We performed the same analysis as described above while ignoring these few spectra, and found that there was no significant change in the extracted transmission spectrum and the results discussed in Section~\ref{sec:results}.

\vspace{1em}
\section{Results and Discussion}
\label{sec:results}

\subsection{The Combined Transmission Spectrum of HAT-P-70~b: Ca\,{\sc ii} Triplet Lines}
\label{sec:combined_transmission_results}

The bottom row of Figure~\ref{fig:transmission_spectrum} shows the combined transmission spectrum around the infrared Ca\,{\sc ii} triplet, representing the average absorption over the full transit. All three Ca\,{\sc ii} lines are clearly resolved and detected at high SNR. In the rest frame of the planet, the three lines should be located at 8498.02, 8542.09, and 8662.14~{\AA} (marked with the blue dashed lines). We took the standard approach to fit Gaussian profiles to each absorption line, and from this, measured their absorption depths ($\delta$), observed (line of sight) velocity offsets of the line centroids from their rest-frame locations, and full width at half maximum (FWHM) -- the results are shown in Table~\ref{tab:results}.

Using Equation (13) from \citet{langeveld2022}, the measured absorption depths ($\delta$) from Table~\ref{tab:results} can be converted to relative atmospheric heights (normalized to $R_{\rm p}$) at which most of the absorption occurs. For the deepest line at 8542.09~{\AA}, with a depth of ($-1.22\pm0.03$)\,\%, the corresponding relative atmospheric height is $0.47\pm0.02$ -- i.e. the radius of the planet at this wavelength appears to be 47\% larger than than the white-light planetary radius due to the absorption occurring within the atmosphere. 

The uncertainty in the mass of HAT-P-70~b, where only an upper limit ($M_{\rm p} < 6.78~M_{\rm J}$) has previously been measured, makes it difficult to quantify if the atmospheric layer probed by Ca\,{\sc ii} is escaping. However, using the upper limit for the planetary mass, the Roche lobe radius approximately extends to $\sim3.9~R_{\rm p}$ \citep{eggleton1983}. The strongest Ca\,{\sc ii} absorption line would only indicate atmospheric escape if the mass of the planet is less than $0.4~M_{\rm J}$. Therefore, it is unlikely that this region of the atmosphere is undergoing atmospheric escape, but it does not rule it out for regions probed by other species that will be investigated in future analyses.

% FWHM in km/s
\begin{table}[]
\centering
\caption{The depths, velocity offsets (deviation from the rest frame wavelengths), and FWHM of the detected Ca\,{\sc ii} triplet lines, measured from Gaussian fits to each line.}
\label{tab:results}
\begin{tabular*}{\columnwidth}{l@{\extracolsep{\fill}}ccc}
\toprule
Ca\,{\sc ii} line & Depth          & Velocity       & FWHM          \\
$\lambda$ ({\AA}) & $\delta$ (\%)  & (km~s$^{-1}$)  & (km~s$^{-1}$)       \\
\midrule
8498.02           & $-0.67\pm0.03$  & $-4.1\pm0.4$  & $19.8\pm1.1$ \\
8542.09           & $-1.22\pm0.03$  & $-3.0\pm0.3$  & $21.1\pm0.7$ \\
8662.14           & $-0.96\pm0.03$  & $-5.1\pm0.4$  & $22.9\pm0.7$ \\
\bottomrule
\end{tabular*}
\end{table}

Whilst this is the first detection of the infrared Ca\,{\sc ii} triplet in the transmission spectrum of HAT-P-70~b, the near ultraviolet Ca\,{\sc ii} H and K lines were previously detected using HARPS-N at the TNG \citep{bello-arufe2022}, with line depths of ($-3.25\pm0.37$)\,\% and ($-4.30\pm0.36$)\,\% respectively. Such strong lines, seen for HAT-P-70~b as well as other UHJs, could be explained by hydrodynamic outflows causing a significant amount of Ca\,{\sc ii} to be transported into the upper atmosphere \citep{yan2019}. We do not analyze the Ca\,{\sc ii} H and K lines in this initial study with GHOST due to the spectrograph focusing issue and the bluest wavelengths suffering from low SNR.

Since efforts were made to ensure that the spectra were combined in the planetary rest frame, any deviations of the measured Gaussian centroids from the rest-frame positions of the Ca\,{\sc ii} triplet lines should originate from physical processes within the atmosphere \citep[although uncertainties in the ephemerides may also contribute, e.g.][]{stangret2024}.
The three Ca\,{\sc ii} lines are observed to be blueshifted by different amounts, between \mbox{$\sim$~3 to 5~km\,s$^{-1}$} along the line of sight, with non-overlapping error ranges. This suggests that strong dayside-to-nightside winds exist at multiple altitudes in the atmosphere of \mbox{HAT-P-70~b}, with a possible vertical change in velocity. These dynamical processes are typical of highly irradiated gas giants \citep[e.g.][]{seidel2020, seidel2021, seidelprinoth2023, seidel2023, langeveld2022, mounzer2022, allart2023}.
Our measured velocities are also comparable to the other individually resolved spectral lines presented by \citet{bello-arufe2022}.

Finally, we verified that the measured velocities are intrinsic to the planet and are not an artefact from \mbox{instrumental} drift. For all observed spectra, we isolated a region containing a number of strong telluric lines close to the Ca\,{\sc ii} triplet, between 8200--8300~{\AA}, and performed a continuum normalization. We then followed the standard high-resolution cross-correlation methodology \citep[e.g.][]{snellen2010, birkby2018}, assigning the first spectrum as the ``template'' to cross-correlate with all other spectra, each time Doppler shifting the template between $\pm$\,50~km\,s$^{-1}$ in steps of 0.05~km\,s$^{-1}$. Gaussian profiles were then fitted to each exposure's cross-correlation function, and the velocities corresponding to the maximum values of the fits were recorded -- see \mbox{Figure~\ref{fig:drift}}. This revealed a drift of $\sim$~0.22~km\,s$^{-1}$ over the $\sim$~5.5~hr sequence of exposures, therefore negligible compared to the observed velocities discussed above, adding validity to the planetary origin.

\subsection{Comparison with Non-Local Thermodynamic Equilibrium Atmospheric Models}
\label{sec:nlte_models}

The atmospheres of UHJs are expected to experience non-local thermodynamic equilibrium (NLTE) effects \citep[e.g.][]{fossati2021}, which can significantly impact the formation of spectral lines, particularly at the high altitudes (low pressures) probed by high-resolution spectroscopy \citep[e.g.][]{young2024, stangret2024}. Moreover, the measured mass of HAT-P-70~b is only an upper limit \citep{zhou2019}, so it is possible for the planet to have a lower mass, which in turn would lower the bulk density and decrease the atmospheric pressure for the same altitude.
We compared the detected Ca\,{\sc ii} triplet lines to NLTE models built using the framework of \citet{young2024}, updated to use the newly released \texttt{Cloudy} version~23.01 \citep{chatzikos2023}. 
Using this framework, we created six NLTE atmospheric models: three with constant Ca abundance ($1\times$ solar) but varying planetary mass (the 6.78~$M_{\rm J}$ upper limit, 2~$M_{\rm J}$, and 1~$M_{\rm J}$); and three with constant planetary mass (2~$M_{\rm J}$) but varying Ca abundance ($1\times$, $10\times$, and $100\times$ solar).

Figure~\ref{fig:NLTE_models} shows a comparison of these NLTE models to the observed  Ca\,{\sc ii} triplet lines in the combined transmission spectrum, with the ``constant abundance'' models in the top panels and the ``constant mass'' on the bottom. To include the observed dayside-to-nightside atmospheric flow within the models, they were each shifted by the measured velocity offset of the Gaussian centroid for each line (listed in Table~\ref{tab:results}), and the $x$-axes are now displayed in velocity space.

For the constant abundance models, it is clear that 6.78~$M_{\rm J}$ is not a realistic mass for HAT-P-70~b, since the observed absorption lines are significantly deeper than simulated. The 2~$M_{\rm J}$ model slightly underestimates and the 1~$M_{\rm J}$ model slightly overestimates the observed line depths. Therefore, it is likely that the true mass of \mbox{HAT-P-70~b} lies somewhere in between these values -- comparable to the mass of other UHJs that also show strong absorption features (e.g. WASP-76~b, WASP-121~b, WASP-189~b). This would also decrease the Roche lobe radius and bring the atmosphere closer to the evaporation regime.
Further support for a lower mass planet is added by the fact that detecting spectral lines in the transmission spectra of high-gravity exoplanets has proven to be difficult and often unsuccessful \citep[e.g.][]{wyttenbach2017, stangret2021, langeveld2022}.
A higher atmospheric temperature could also produce stronger absorption lines; however, the pressure-temperature profiles generated with these models already factor in a temperature that extends several thousands of degrees hotter at lower pressures than the equilibrium temperature.

These inferences are supported by high-resolution atmospheric retrievals that were performed on the transit observations of HAT-P-70~b with HARPS-N; with $\log(g)$ as a free parameter in the retrievals, \citet{gandhi2023} constrained the planetary mass to $1.66\pm0.20~M_{\rm J}$ (agreeing with our constraint from the NLTE models), albeit with the caveat that degeneracies in the retrieved parameters exist since the scale height of the atmosphere can also be affected by the temperature, the mean molecular weight of the atmosphere (which is assumed), and recombination of H to H$_2$.

For the constant mass models, we fixed the mass to 2~$M_{\rm J}$ to investigate if a super-solar abundance could deepen the absorption and provide a closer match to the observations. However, as seen in the bottom panels of Figure~\ref{fig:NLTE_models}, the wings of the absorption lines for the $10\times$ and $100\times$ solar models are much broader than perceived in the observations, with the $1\times$ solar abundance model more closely matching the line profiles in the data. Therefore, it is likely that the abundance of Ca in the atmosphere of HAT-P-70~b is approximately solar or subsolar.
Incorporation of high-resolution atmospheric retrievals \citep[e.g.][]{gandhi2023, pelletier2023} with these GHOST data will provide clarity on this matter and will be explored in future work. Such retrieval frameworks extract the net atmospheric abundances utilizing the cross-correlation technique, and an atmospheric model of Ca\,{\sc ii} may include additional weaker absorbing lines over GHOST's spectral range. Since the atmospheric pressure probed by absorption lines varies depending on the line depth \citep[e.g.][]{young2024, kesseli2024}, retrievals may probe higher-pressure (deeper) regions of the atmosphere, where the abundance of Ca\,{\sc ii} may be different than the lower-pressure region probed by the triplet lines \citep[as is expected based on the abundance profiles for multiple species in UHJ atmospheres, e.g.][]{fossati2021}. Therefore, comparison of retrievals over different wavelength ranges provided by GHOST, or by selecting only the absorption lines within a certain range of depths, may provide further insight into the atmospheric structure and abundance of Ca\,{\sc ii} at different altitudes.

\subsection{Time-Resolved Transmission Spectroscopy of the Individual Ca\,{\sc ii} Triplet Lines}
\label{sec:time-Resolved_results}

The temporal variation of the atmospheric signals can be assessed by dividing the in-transit residual spectra (middle panel of Figure~\ref{fig:transmission_spectrum}) into multiple bins, then combining the spectra within each phase bin to measure parameters of the Ca\,{\sc ii} lines from subsequent Gaussian fits (as performed in Section~\ref{sec:combined_transmission_results}).
To achieve this, phases from the transit ingress until \mbox{$\phi\sim-0.01$} (where the RM and atmospheric signals overlap) were ignored, and the remaining data were divided equally into six groups of 10 spectra. The choice of six phase bins ensured that there are enough points to assess the temporal variation (with at least two bins before the mid-transit), and sufficient SNR within each bin to robustly resolve the Ca\,{\sc ii} lines and measure their parameters with Gaussian fits.

\begingroup
\begin{figure*}[p]
    \centering
    \includegraphics[width=\textwidth]{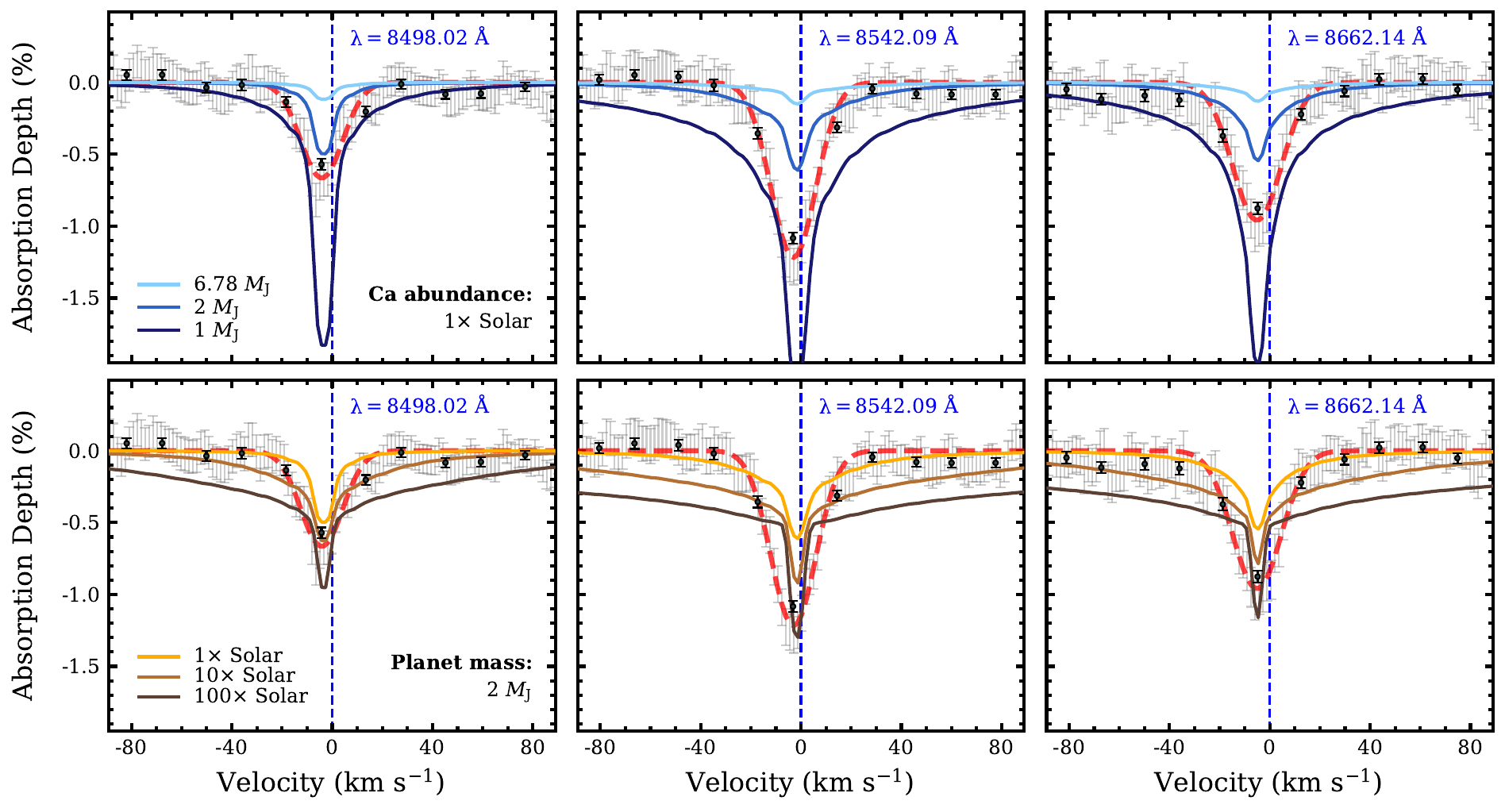}
    \vskip-2mm
    \caption{The Ca\,{\sc ii} triplet lines in the GHOST transmission spectrum of HAT-P-70~b compared to NLTE models generated for different planetary masses and Ca abundances. The same data from Figure~\ref{fig:transmission_spectrum} is displayed (full-resolution in grey, binned 10$\times$ in black, Gaussian fit in red), now in velocity space with the rest-frame line positions at 0~km~s$^{-1}$. 
    \textit{Top row}: NLTE models for constant abundance ($1\times$~solar) and three different planetary masses (6.78~$M_{\rm J}$, 2~$M_{\rm J}$, and 1~$M_{\rm J}$); the cores of the absorption lines become wider and deeper for decreasing masses.
    \textit{Bottom row}: NLTE models for constant mass (2~$M_{\rm J}$) and three different Ca abundances \mbox{($1\times$, $10\times$} and $100\times$~solar); the line cores become deeper and the wings become wider for higher abundances.}
    \label{fig:NLTE_models}

    \vspace{1em}
    \includegraphics[width=0.85\textwidth]{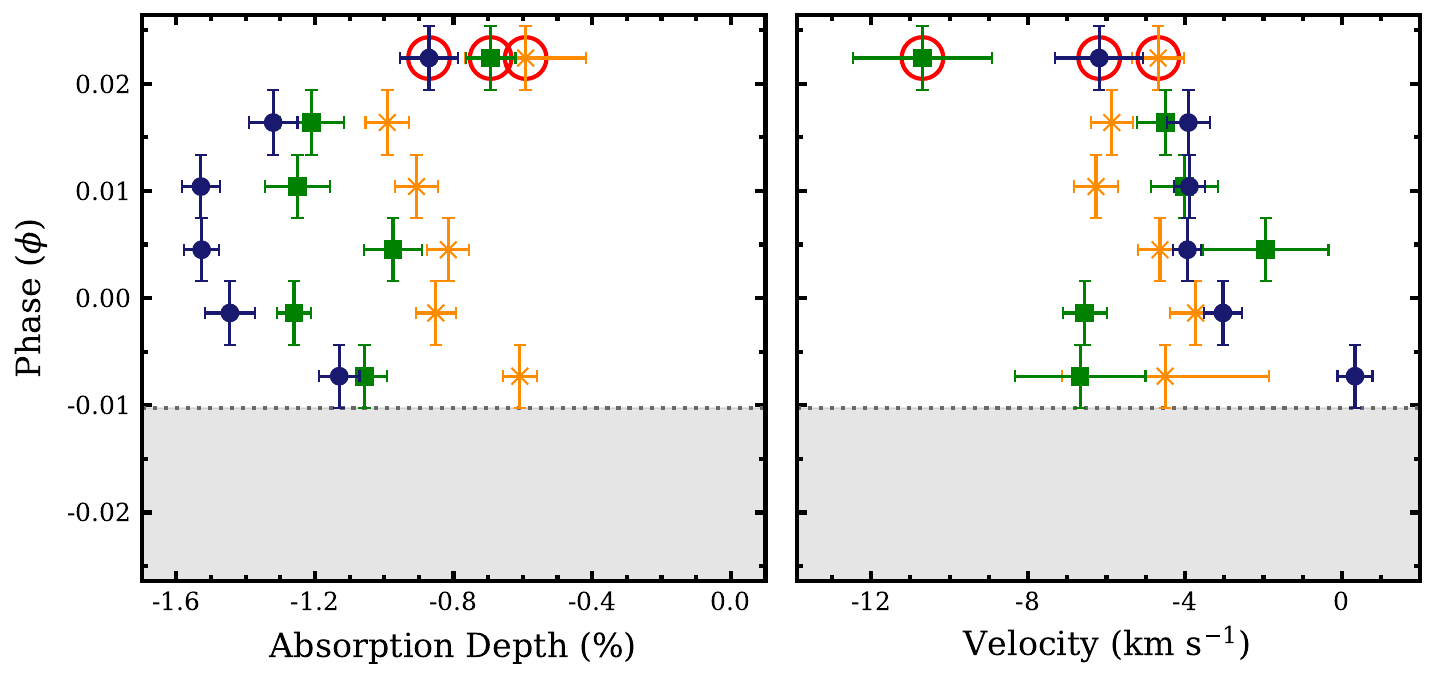}
    \vskip-1mm
    \caption{Variation of the measured absorption depths (\textit{left}) and observed velocity offsets (\textit{right}) of the three Ca\,{\sc ii} triplet lines at several phases throughout the transit of HAT-P-70~b. The phases where the RM features overlapped with the planetary absorption features were ignored (grey shaded region). The remaining in-transit residual spectra were binned into six groups, each containing 10 spectra that were subsequently combined with a weighted average. Each point is measured from Gaussian fits to the Ca\,{\sc ii} absorption lines with rest-frame positions at 8498.02 (orange crosses), 8542.09 (dark blue circles), and 8662.14~{\AA} (green squares). 
    Measurements for the egress phases (outlined with red circles) that probe mostly the trailing limb (evening terminator) show a distinct difference from the other phases -- i.e. stronger blueshifted velocities and weaker absorption depths.}
    \vspace{1em}
    \label{fig:phase_trends}
\end{figure*}
\endgroup

Figure~\ref{fig:phase_trends} shows the variation of the measured absorption depths and velocity offsets of the three Ca\,{\sc ii} lines (orange, dark blue, and green points) across the six phase bins. The points circled in red highlight the phase bin containing data obtained during the egress phases when the planet is partially transiting, thus it is mostly the trailing (evening) limb that contributes to these measurements. The phase bin closest to the masked-out region may not fully contain the atmospheric signal due to the proximity of nearby masked pixels, but we include them for completeness.

In general, the absorption depths of the Ca\,{\sc ii} lines stay between a range of around $-1.6$\,\% (for the strongest line) to $-0.9$\,\% (for the weakest line) across the transit, except for the egress phase bin where all three absorption lines become weaker. This is expected given that the transmission through the leading limb starts to disappear during the egress.
Next, the observed line-of-sight velocities are distributed around a value of $\sim-4$~km\,s$^{-1}$ throughout the transit, except again for the egress phase bin where they are more blueshifted on average. The extra blueshift may be attributed to: (1) stronger dayside-to-nightside winds at the evening terminator driven by an eastwards advection of the hottest region of the atmosphere from the substellar point (the strength of which may depend on atmospheric processes such as hydrogen dissociation/recombination, internal heat flux, or magnetic drag, e.g. \citealt{tan2019, komacek2022, beltz2022, beltz2023, beltz2024}); and/or (2) the Doppler shift of the atmospheric signals from only the blueshifted side of the planet due to the tidally locked planetary rotation \citep{wardenier2021, wardenier2022}. The rotational velocity of the atmosphere assuming tidally locked rotation ranges from zero (at the poles) to a maximum (at the equator) of $\sim$~3.5~km\,s$^{-1}$ for the bulk planetary radius, or $\sim$~5.1~km\,s$^{-1}$ for the extended atmospheric radius probed by the deepest Ca\,{\sc ii} line -- these values roughly match the average velocity difference at egress seen in Figure~\ref{fig:phase_trends} (compared to the rest of the transit), however, we cannot rule out stronger evening terminator winds without more robust measurements across multiple species.
Similar detections have been made for other highly irradiated gas giant exoplanets \citep[e.g.][]{louden2015, ehrenreich2020}.
Therefore, not only can these GHOST data constrain net dayside-to-nightside atmospheric wind speeds from the Ca\,{\sc ii} lines in the combined transmission spectrum, but the time-resolved measurements now provide independent spatial information about the different limbs.

Time-resolved high-resolution transmission spectroscopy of the UHJ WASP-76~b previously revealed an asymmetric shape to the cross-correlated Fe\,{\sc i} absorption trail throughout the transit, where the signal gradually became more blueshifted until around the mid-transit, after which it remained at a constant velocity \citep{ehrenreich2020}. This was explained by Fe\,{\sc i} being localized to a region closer to the evening terminator than the morning, and the rotation of the planet aligning this region such that absorption occurs only within the evening terminator during the entire second half of the transit (thus at a constant velocity). 
These asymmetries were also confirmed to be present at different altitudes within the atmosphere of WASP-76~b from an analysis of the cross-correlated signal for weak (low altitude), medium, and strong (high altitude) Fe\,{\sc i} lines separately \citep{kesseli2024} -- tentative signs of stronger blueshifts at lower altitudes were also reported, in agreement with inferences made from the analysis of the Na doublet \citep{seidel2021}. In contrast, these time-resolved detections of the individual Ca\,{\sc ii} lines for HAT-P-70~b show that the velocity values are overall more blueshifted only at the egress phases, suggesting that Ca\,{\sc ii} is at least partially visible through the leading limb at all phases except when it starts to egress, thus it is distributed across the entire terminator. The signals nearest the masked-out region may suggest some asymmetry in the strength of the lines as well as their Doppler shift; however, we cannot make a significant conclusion about this without a robust removal of the RM signals near ingress.

\section{Conclusion}
\label{sec:conclusions}

This work demonstrates the capabilities of the new Gemini High-resolution Optical SpecTrograph (GHOST) at the Gemini South Observatory for investigating the 3D nature of exoplanet atmospheres. In this first transmission spectroscopy study with the high-resolution mode GHOST, we observed one transit of the ultra-hot Jupiter HAT-P-70~b and made new detections of the infrared Ca\,{\sc ii} triplet absorption lines, complementing previous detections of the near ultraviolet Ca\,{\sc ii} H and K lines \citep{bello-arufe2022}. From Gaussian fits to the three absorption lines, we inferred: (1) that Ca\,{\sc ii} in the planetary atmosphere extends to the upper altitudes of the planetary atmosphere at a height of at least 0.47 times the bulk planetary radius $R_{\rm p}$; and (2) that the three lines are blueshifted, varying between $\sim$~3~to~5~km\,s$^{-1}$ along the line of sight, which is indicative of strong dayside-to-nightside winds with a possible velocity gradient between different altitudes. 
Comparing the combined transmission spectrum with NLTE atmospheric models generated for \mbox{HAT-P-70~b} revealed that the planetary mass is likely to be around 1--2~$M_{\rm J}$ (much lighter than the 6.78~$M_{\rm J}$ upper limit measured from radial velocities), and that the abundance of Ca within the atmosphere is solar or subsolar.

Most notably, capitalizing on the high SNR achieved by GHOST and Gemini South, we showed that the individual Ca\,{\sc ii} absorption lines can be temporally resolved to analyze how the absorption through the terminator changes throughout the transit, allowing different planetary longitudes to be probed (due to rotation of the tidally locked planet). During the transit egress, where the signals mostly originate from the trailing limb (evening terminator), the absorption strength decreased (due to the loss of signal from the leading limb) and the velocities became more blueshifted on average (due to the planet rotating towards the observer at the trailing limb and stronger winds driven by the hotspot being offset eastwards from the substellar point).
Besides the egress phases, we saw no clear evidence of asymmetries in the velocity measurements, suggesting that Ca\,{\sc ii} is distributed at least partially on both the morning and evening terminators. However, this needs to be confirmed with a robust correction of the spurious signals produced by the Rossiter-McLaughlin effect that obscure the atmospheric signals near the planetary ingress.

Time-resolved detections are paramount for advancing our understanding of the complex chemical and dynamical processes in UHJ atmospheres, and they provide crucial contributions towards inferences made with high-resolution atmospheric retrievals and general circulation models. However, the high SNR required for time-resolved detections of individual spectral lines (i.e. without relying on the cross-correlation technique) is only feasible with the latest state-of-the-art high-resolution spectrographs at 8--10\,m class telescopes. Therefore, this demonstration places GHOST among the few instruments that can achieve these goals.

%% Also note that the akcnowlodgment environment does not support long amounts of text. If you have a lot of people and institutions to acknowledge, do not use this command. Instead, create a new \section{Acknowledgments}.
\section*{Acknowledgments}
%\begin{acknowledgments}
We thank the anonymous referee for their constructive and thoughtful comments, which helped to improve this manuscript.

The authors thank the observers at Gemini South for the data acquisition (Venu Kalari, Lindsay Magill, Daniel May, and Henrique Reggiani), and  Jeong-Eun Heo (program contact scientist with EKD).

Based on observations obtained under Program ID GS-2023B-FT-105 at the international Gemini Observatory, a program of NSF NOIRLab, which is managed by the Association of Universities for Research in Astronomy (AURA) under a cooperative agreement with the U.S. National Science Foundation on behalf of the Gemini Observatory partnership: the U.S. National Science Foundation (United States), National Research Council (Canada), Agencia Nacional de Investigaci\'{o}n y Desarrollo (Chile), Ministerio de Ciencia, Tecnolog\'{i}a e Innovaci\'{o}n (Argentina), Minist\'{e}rio da Ci\^{e}ncia, Tecnologia, Inova\c{c}\~{o}es e Comunica\c{c}\~{o}es (Brazil), and Korea Astronomy and Space Science Institute (Republic of Korea).
Data processed using DRAGONS (Data Reduction for Astronomy from Gemini Observatory North and South).
% https://noirlab.edu/science/about/scientific-acknowledgments#ack-gemini

GHOST was built by a collaboration between Australian Astronomical Optics at Macquarie University, National Research Council Herzberg of Canada, and the Australian National University. The instrument scientist is Dr. Alan McConnachie at NRC, and the instrument team is also led by Dr. Gordon Robertson (at AAO), and Dr. Michael Ireland (at ANU).

EKD acknowledges the support of the Natural Sciences and Engineering Research Council of Canada (NSERC), funding reference No. 568281-2022.

EdM and MY acknowledge support from the Science and Technology Facilities Council (STFC) award ST/X00094X/1.

JDT acknowledges funding support by the TESS Guest Investigator Program G06165. 

% \end{acknowledgments}


\vspace{2em}
\facilities{Gemini:South (GHOST)}


\vspace{2em}
\software{
\texttt{GHOSTDR} \citep{ireland2018, hayes2022}, 
\texttt{DRAGONS} \citep{labrie2023}, 
\texttt{molecfit} \citep{smette2015, kausch2015}, 
\texttt{Cloudy} \citep{chatzikos2023}, 
\texttt{Astropy} \citep{astropy2013, astropy2018, astropy2022}, 
\texttt{NumPy} \citep{harris2020}, 
\texttt{SciPy} \citep{virtanen2020}, 
\texttt{Matplotlib} \citep{hunter2007}
}



\vspace{1em}
\appendix
\vspace{-2em}
\restartappendixnumbering

\section{System Parameters}
The parameters adopted during the extraction and analysis of the HAT-P-70~b transmission spectrum are listed in Table~\ref{tab:parameters}. There is only an upper limit for the planetary mass (and therefore planetary RV semi-amplitude).

\begingroup
\renewcommand{\arraystretch}{1.25} % Vertical (row) spacing, default value: 1
\begin{table*}
    \centering
    \caption{Stellar, planetary, and orbital parameters adopted in this work for the HAT-P-70 system. 
    Derived parameters:
    (${\dagger}$)~Calculated using \mbox{$K_{\text{p}} = -K_{\ast}(M_{\ast}/M_{\text{p}})$}, but due to the constraints on $K_\ast$ and $M_{\text{p}}$, this value is derived using the upper limit of both parameters; 
    % $^{(\ddagger)}$~Derived using $T_{\text{eq}} = T_{\text{eff}}\left[\frac{R_\ast^2}{2a^2}(1-f_\text{r})(1-A_\text{B})\right]^{\frac{1}{4}}$, assuming uniform heat redistribution ($f_\text{r} = \frac{1}{2}$), zero bond albedo, and a circular orbit.
    (${\ddagger}$)~Calculated using $T_{\text{eq}} = T_{\text{eff}}\left[(1-f_{\text{r}})(1-A_{\text{B}})R_\ast^2/(2a^2)\right]^{1/4}$, assuming uniform heat redistribution ($f_{\text{r}} = 0.5$), zero bond albedo, and a circular orbit.}
    \begin{tabular*}{\textwidth}{l@{\extracolsep{\fill}}llll}
    \toprule
    Parameter                     & Symbol           & Value                            & Unit                     & Reference           \\
    \midrule
    \textbf{Star}                 &                  &                                  &                          &                     \\
    Stellar Mass                  & $M_\ast$         & 1.890 $^{+0.010}_{-0.013}$       & $M_{\sun}$               & \citet{zhou2019}    \\
    Stellar Radius                & $R_\ast$         & 1.858 $^{+0.119}_{-0.091}$       & $R_{\sun}$               & \citet{zhou2019}    \\
    Stellar RV Semi-amplitude     & $K_\ast$         & $< 649$                          & m s$^{-1}$               & \citet{zhou2019}    \\
    Effective Temperature         & $T_{\text{eff}}$ & 8450 $^{+540}_{-690}$            & K                        & \citet{zhou2019}    \\
    Projected Rotational Velocity &$v\sin{i}$        & 99.85 $^{+0.64}_{-0.61}$         & km s$^{-1}$              & \citet{zhou2019}    \\
    Surface Gravity               & $\log{g}$        & 4.181 $^{+0.055}_{-0.063}$       & $\log_{10}$(cm s$^{-2}$) & \citet{zhou2019}    \\
    Metallicity                   & [Fe/H]           & -0.059 $^{+0.075}_{-0.088}$      & dex                      & \citet{zhou2019}    \\
    \addlinespace
    \textbf{Planet}               &                  &                                  &                          &                     \\
    Planetary Mass                & $M_{\text{p}}$   & $< 6.78$                         & $M_{\text{J}}$           & \citet{zhou2019}    \\
    Planetary Radius              & $R_{\text{p}}$   & 1.87 $^{+0.15}_{-0.10}$          & $R_{\text{J}}$           & \citet{zhou2019}    \\
    Planetary RV Semi-amplitude   & $K_{\text{p}}$   & $-190$                           & km s$^{-1}$              & Derived $^\dagger$  \\
    Equilibrium Temperature       & $T_{\text{eq}}$  & 2550 $^{+180}_{-220}$            & K                        & Derived $^\ddagger$ \\
    Orbital Inclination           & $i_{\text{p}}$   & 96.50 $^{+1.42}_{-0.91}$         & deg                      & \citet{zhou2019}    \\
    Sky Projected Obliquity       & $\lambda$        & 113.1 $^{+5.1}_{-3.4}$           & deg                      & \citet{zhou2019}    \\
    \addlinespace
    \textbf{System}               &                  &                                  &                          &                     \\
    Period                        & $P$              & 2.744320 $^{+0.000001}_{-0.000001}$    & days               & \citet{ivshina2022} \\
    Mid-transit Time              & $T_{\text{c}}$   & 2459175.05277 $^{+0.00017}_{-0.00017}$ & BJD                & \citet{ivshina2022} \\
    Transit Duration              & $T_{14}$         & 0.145 $^{+0.003}_{-0.002}$       & days                     & \citet{zhou2019}    \\
    Semi-major Axis               & $a$              & 0.04739 $^{+0.00031}_{-0.00106}$ & au                       & \citet{zhou2019}    \\
    Systemic Velocity             & $v_{\text{sys}}$ & 25.260 $^{+0.110}_{-0.110}$      & km s$^{-1}$              & \citet{zhou2019}    \\
    \addlinespace
    \bottomrule
    \end{tabular*}
    \label{tab:parameters}
\end{table*}
\endgroup

\section{Instrumental Drift}
\begin{figure*}
    \centering
    \includegraphics[width=0.85\textwidth]{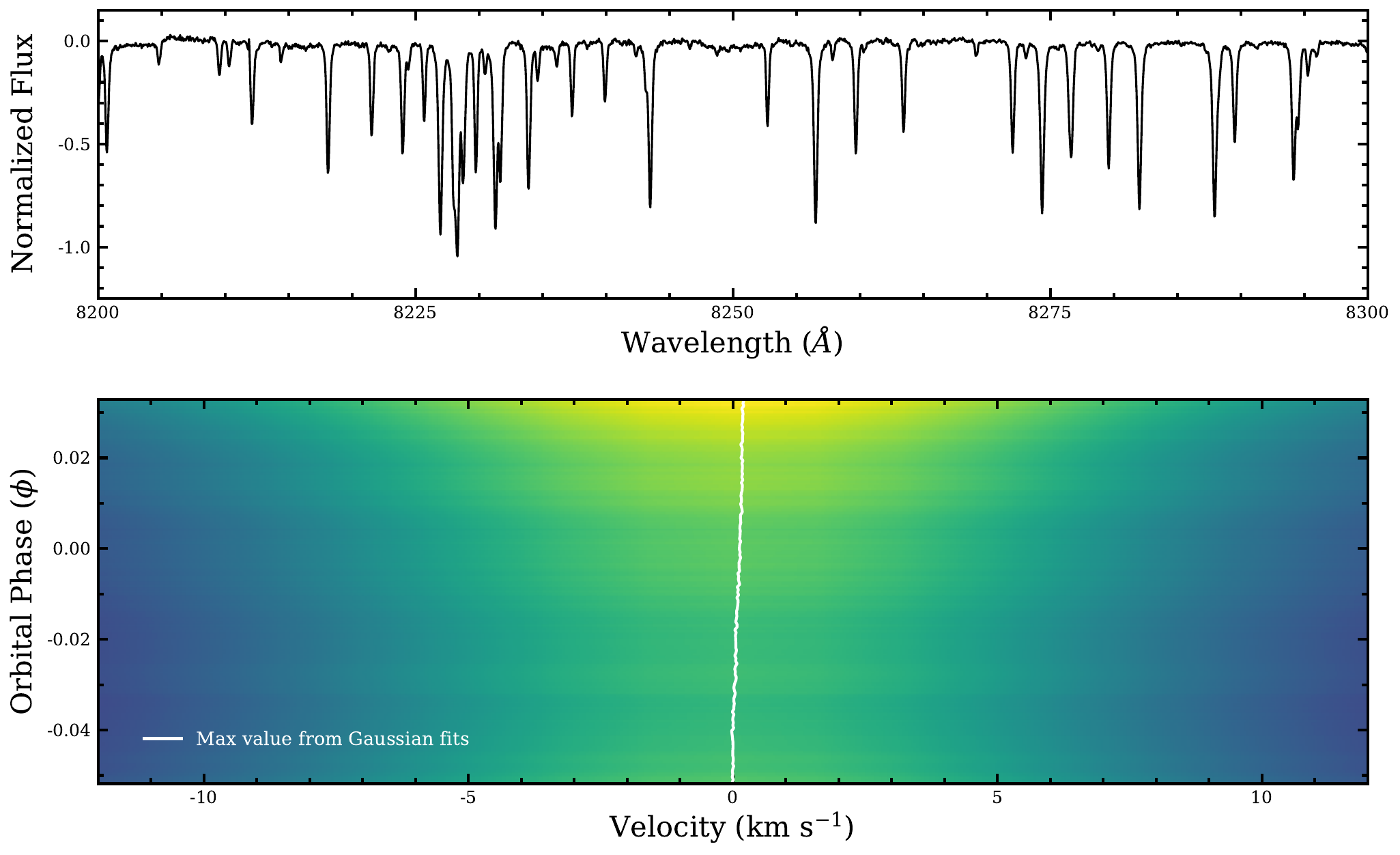}
    \caption{The instrumental drift of GHOST for the duration of these observations. \textit{Top panel:} isolated region of the spectrum (continuum normalized) containing a number of strong telluric lines that was used as a template for the cross-correlation analysis. \textit{Bottom panel:} result after cross-correlating each observed spectrum with the template (Doppler shifted between $\pm$\,50~km\,s$^{-1}$ in steps of 0.05~km\,s$^{-1}$), displayed in phase--velocity space. Brighter colours indicate a stronger correlation. The maximum value of each cross-correlation function (each row), measured with Gaussian fits, is marked by the white line; the total drift is $\sim$~0.22~km\,s$^{-1}$ over the sequence of exposures.}
    \vspace{3em}
    \label{fig:drift}
\end{figure*}

The instrumental drift of GHOST over the $\sim$~5.5~hr sequence of exposures was measured as described in Section~\ref{sec:combined_transmission_results} and illustrated in Figure~\ref{fig:drift}. The measured drift is negligible compared to observed velocity offsets of the spectral lines from the atmosphere of HAT-P-70~b.

\bibliography{manuscript}{}
\bibliographystyle{aasjournal}



\end{document}